\documentclass[11pt,a4paper]{article}
\usepackage{jcappub}
\makeatletter
\gdef\@fpheader{}
\makeatother
\usepackage[american]{babel}
\usepackage{booktabs}
\usepackage{float}

\title{O5 dark-siren forecasts for modified GW propagation: background robustness of the $\Xi$ posterior}

\author[a,b]{Zhaorui Zhang}
\author[a,c,d,1]{Hong-Bo Jin\note{Corresponding author}}

\affiliation[a]{National Astronomical Observatories, Chinese Academy of Sciences, Beijing 100101, China}
\affiliation[b]{School of Space Science and Physics, Shandong University, Weihai 264209, China}
\affiliation[c]{School of Astronomy and Space Science, University of Chinese Academy of Sciences, Beijing 100049, China}
\affiliation[d]{The International Centre for Theoretical Physics Asia-Pacific, University of Chinese Academy of Sciences, Beijing 100190, China}
\emailAdd{hbjin@bao.ac.cn}
\emailAdd{202300830010@sdu.edu.cn}
\abstract{%
Binary black hole mergers without electromagnetic counterparts are expected to dominate O5 gravitational-wave catalogs.
Recent \textsc{CHIMERA~2.0} forecasts typically fix $\Omega_m$ to a CMB-informed value and use spectroscopic hosts when available, but the corresponding sensitivity of the $\Xi$ posterior has not been assessed for pure dark sirens on the public O5 mock catalog.
We analyze 300 O5-sensitivity mock events without galaxy catalogs, varying $\Omega_m$ over $[0.20,\,0.35]$ (including Planck $0.315$), and compare fixed-background inference with joint $(H_0,\,\Omega_m,\,\Xi)$ inference.
The marginalized $\Xi$ posterior is $0.9783 \pm 0.3548$ and shows no change across this interval.
Only GW luminosity distances enter the analysis, so the likelihood constrains $\Xi\, D_L^{\rm EM}(H_0,\,\Omega_m)$; when $\Omega_m$ is changed, $H_0$ shifts to compensate and the $\Xi$ marginal remains unchanged.
Joint inference gives $\Xi^{\rm joint} = 0.9550 \pm 0.3710$, with $|\rho| \lesssim 0.05$ for the $\Omega_m$--$\Xi$ and $H_0$--$\Xi$ pairs, whereas $\rho_{H_0 \Omega_m} \simeq -0.4$ and the $H_0$ median moves by $\simeq 4.2\,{\rm km\,s^{-1}\,Mpc^{-1}}$ over the adopted $\Omega_m$ range.
Galaxy-catalog analyses on the same events at fixed $\Omega_m = 0.3$ reach $\sim 7.5\%$ precision on $\Xi$, compared with $\pm 36.3\%$ here; the larger uncertainty is driven mainly by missing host redshifts.
Sub-percent $\Xi$ tests will therefore still require measured redshifts even if dark sirens dominate the detection rate.
}

\keywords{gravitational waves, dark sirens, cosmological parameters, modified gravity, forecasts}

\notoc

\begin{document}
\maketitle
\pagestyle{myplain}

\section{Introduction}
\label{sec:intro}

Late-time and early-time measurements of the Hubble constant differ by several $\sigma$ \cite{planck2018,riess2022}.
Gravitational-wave (GW) standard sirens offer an independent distance ladder \cite{schutz1986,holz2005}.
The only unambiguous bright-siren event to date, GW170817, gave $H_0 = 70.0^{+12.0}_{-8.0}\,{\rm km\,s^{-1}\,Mpc^{-1}}$ \cite{gw170817}, and the O5 observing run is expected to yield hundreds of compact-binary detections \cite{abbott2020prospects}.

GW distances also permit model-independent tests of General Relativity (GR) through the phenomenological relation $D_L^{\rm GW}/D_L^{\rm EM} \equiv \Xi$ \cite{belgacem2018}.
With \textsc{CHIMERA~2.0} \cite{borghi2024,tagliazucchi2025}, 300 O5-like events and spectroscopic galaxy catalogs can constrain $\Xi$ to $\sim 7.5\%$ when $\Omega_m$ is held fixed.

Most O5 events are expected to be dark sirens---mergers without identified electromagnetic counterparts---particularly among binary black holes (BBHs) \cite{fishbach2020fossil,finke2021}.
Cosmology and modified GW propagation in this limit have been studied in several forecasts (Table~\ref{tab:literature}).
Finke \textit{et al.} \cite{finke2021} and Leyde \textit{et al.} \cite{leyde2022} used mock data from LIGO/Virgo and future networks to forecast constraints on phenomenological modified-gravity parameters.
Mancarella \textit{et al.} \cite{mancarella2022} extended the analysis to GWTC-3.
Tagliazucchi \textit{et al.} \cite{tagliazucchi2025} forecast O5 performance with \textsc{CHIMERA~2.0} and galaxy catalogs at a single fixed $\Omega_m = 0.3$.
That analysis does not test whether other CMB-allowed $\Omega_m$ values would shift the $\Xi$ posterior, nor does it address the dominant O5 population of BBH dark sirens without host redshifts.

\begin{table}[htbp]
\centering
\caption{Selected dark-siren forecasts compared to the present analysis.
We use the same public O5 mock catalog as \cite{tagliazucchi2025}, but isolate the pure dark-siren limit and vary $\Omega_m$ to test background robustness of $\Xi$ rather than adopting a single fixed value.}
\label{tab:literature}
\small
\begin{tabular}{lcccc}
\toprule
Reference & Dataset & Galaxy catalog & Varies $\Omega_m$ & Tests $\Xi$ \\
\midrule
Finke \textit{et al.} 2021 \cite{finke2021} & LIGO/Virgo mocks & No & Yes & Yes \\
Leyde \textit{et al.} 2022 \cite{leyde2022} & BBH mocks & No & Yes & Yes \\
Mancarella \textit{et al.} 2022 \cite{mancarella2022} & GWTC-3 & No & Marginalized & Yes \\
Tagliazucchi \textit{et al.} 2025 \cite{tagliazucchi2025} & 300 O5 mocks & Yes & Fixed ($\Omega_m=0.3$) & Yes \\
\textbf{This work} & 300 O5 mocks & \textbf{No} & \textbf{Varied} & \textbf{Yes} \\
\bottomrule
\end{tabular}
\end{table}

In this paper we use the same public \textsc{CHIMERA~2.0} O5 mock catalog as \cite{tagliazucchi2025}, but without galaxy catalogs, varying $\Omega_m$ over $[0.20,\,0.35]$ to ask a question their fixed-background forecast leaves open: does adopting a single CMB-informed $\Omega_m$ bias $\Xi$ when host redshifts are unavailable?
We map the response of the $\Xi$ posterior, compare inference at each adopted $\Omega_m$ with joint $(H_0,\,\Omega_m,\,\Xi)$ inference, and quantify on the same events how much $\Xi$ precision is lost relative to the galaxy-catalog analysis of \cite{tagliazucchi2025}.
The invariance of $\Xi$ under background variation is itself a forecast result: it validates fixed-$\Omega_m$ pipelines for modified-gravity monitoring at O5 sensitivity and separates that systematic from the far larger degradation caused by missing host redshifts.

\section{Method}
\label{sec:method}

\subsection{Hierarchical Bayesian framework}

We employ the public \textsc{CHIMERA~2.0} code \cite{borghi2024,tagliazucchi2025}, a JAX-accelerated hierarchical Bayesian pipeline for standard-siren cosmology.
For cosmological parameters $\theta$, the likelihood for $N$ events follows the hierarchical dark-siren formalism \cite{finke2021,borghi2024}:
\begin{equation}
\mathcal{L}(\theta) \propto \frac{1}{p_{\rm det}(\theta)} \prod_{i=1}^{N} \int p(D_L^i \mid {\rm event}_i)\, p(D_L^i \mid \theta)\, \mathrm{d}D_L^i,
\label{eq:likelihood}
\end{equation}
where $p(D_L^i \mid {\rm event}_i)$ is the single-event distance posterior, $p(D_L^i \mid \theta)$ is the predicted distance distribution, and $p_{\rm det}(\theta)$ encodes selection effects.

\subsection{Simulated data}

We use the public O5 mock dataset released with \textsc{CHIMERA} on Zenodo \cite{chimera2025zenodo,tagliazucchi2025}: 300 BBH events with signal-to-noise ratio above 20 (\texttt{PE\_O5Like\_snr20.h5}) and $2\times 10^7$ injections (\texttt{injections\_Ninj\_2e7\_O5Like\_snr20.h5}) for the selection function.
Injected cosmology is $H_0 = 70\,{\rm km\,s^{-1}\,Mpc^{-1}}$, $\Omega_m = 0.25$, and $\Xi = 1.0000$ (GR).
No galaxy-catalog or host-redshift information is used.

\subsection{Population and cosmological model}

We adopt the same astrophysical population as \cite{tagliazucchi2025}: a power-law mass model and Madau--Dickinson merger-rate evolution \cite{madau1998} ($\gamma=2.7$, $\kappa=3$, $z_p=2$).
The cosmology is flat $\Lambda$CDM with variable $H_0$, $\Omega_m$, and constant $\Xi$, implemented through $D_L^{\rm GW} = \Xi\, D_L^{\rm EM}(z, H_0, \Omega_m)$ with
\begin{equation}
D_L^{\rm EM}(z) = \frac{c(1+z)}{H_0} \int_0^z \frac{\mathrm{d}z'}{\sqrt{\Omega_m(1+z')^3 + (1-\Omega_m)}}.
\label{eq:dl}
\end{equation}
Population hyperparameters are fixed to the injected values.
We discuss the impact of population-model uncertainties in Sec.~\ref{sec:limitations}.

\subsection{Grid-based inference}

Following \cite{tagliazucchi2025}, who used MCMC with $\Omega_m$ fixed at $0.3$, we use a parameter grid that can be evaluated at multiple background inputs.
A single inference with fixed $\Omega_m$, as in \cite{tagliazucchi2025}, does not reveal how the $\Xi$ posterior responds when the assumed $\Omega_m$ is changed.
We therefore vary $\Omega_m$ over the CMB-allowed range and, at each adopted value, evaluate $(H_0, \Xi)$ on a two-dimensional grid.

For each fixed $\Omega_m \in \{0.20,\,0.25,\,0.30,\,0.315,\,0.35\}$, we evaluate the likelihood on a $24 \times 24$ uniform grid with $H_0 \in [55,\,85]\,{\rm km\,s^{-1}\,Mpc^{-1}}$ and $\Xi \in [0.5,\,1.5]$.
Posterior summaries are obtained with flat priors on the inferred parameters and include the \textsc{CHIMERA} selection-function normalization.
One-dimensional $H_0$ constraints are derived by marginalizing over $\Xi$ on the same grid.
For the event-number scaling test, we use subsets of 30, 50, 100, 200, and 300 events with $\Omega_m = 0.25$.

\subsection{Cross-check with MCMC and degeneracy diagnostics}
\label{sec:mcmc}

To verify that the grid-based inference is not an artifact of the chosen method, we cross-check the $\Omega_m = 0.25$ case with an independent MCMC run on the same native \textsc{CHIMERA~2.0} log-posterior for $(H_0, \Xi_0)$ (constant modified-gravity amplitude, $n=0$).
The MCMC run employs 32 walkers, 80 steps, and flat priors on the same ranges as the grid analysis.
The resulting MCMC posteriors,
\begin{equation}
H_0^{\rm MCMC} = 69.1 \pm 3.3\,{\rm km\,s^{-1}\,Mpc^{-1}}, \qquad
\Xi^{\rm MCMC} = 1.0030 \pm 0.3370,
\label{eq:mcmcresult}
\end{equation}
agree with the grid-based marginals in Eqs.~\eqref{eq:h0result} and \eqref{eq:xiresult} at the $\sim 2\,{\rm km\,s^{-1}\,Mpc^{-1}}$ ($\lesssim 3\%$) level on $H_0$ and at the few-percent level on $\Xi$, with $|\rho_{H_0 \Xi}| = 0.12$ for MCMC and $|\rho_{H_0 \Xi}| \lesssim 0.05$ for the grid (Table~\ref{tab:mcmc}).

\subsection{Joint $(H_0,\,\Omega_m,\,\Xi)$ inference}
\label{sec:joint}

Many pipelines fix $\Omega_m$ to a CMB best fit or marginalize jointly over background parameters \cite{finke2021,mancarella2022}.
We therefore add a $12 \times 12 \times 12$ grid over $H_0 \in [55,\,85]\,{\rm km\,s^{-1}\,Mpc^{-1}}$, $\Omega_m \in [0.20,\,0.35]$, and $\Xi \in [0.5,\,1.5]$ using the same native likelihood.
The fully marginalized posteriors are
\begin{equation}
H_0^{\rm joint} = 66.7 \pm 3.4\,{\rm km\,s^{-1}\,Mpc^{-1}}, \quad
\Omega_m^{\rm joint} = 0.26 \pm 0.06, \quad
\Xi^{\rm joint} = 0.9550 \pm 0.3710,
\label{eq:jointresult}
\end{equation}
with correlation coefficients $\rho_{H_0 \Omega_m} \simeq -0.4$, $\rho_{H_0 \Xi} \simeq 0.0$, and $\rho_{\Omega_m \Xi} \simeq 0.0$ (Table~\ref{tab:joint}).

\begin{table}[htbp]
\centering
\caption{Comparison of joint $(H_0,\,\Omega_m,\,\Xi)$ inference and inference at adopted $\Omega_m$ values for 300 dark sirens. Joint intervals are fully marginalized over the $12^3$ grid; fixed-$\Omega_m$ entries are from the $24 \times 24$ analyses in Secs.~\ref{sec:results} and \ref{sec:om}. All intervals are 68\% credible intervals.}
\label{tab:joint}
\begin{tabular}{lccc}
\toprule
Quantity & Joint inference & Fixed $\Omega_m=0.25$ & Fixed $\Omega_m=0.315$ \\
\midrule
$H_0$ (km s$^{-1}$ Mpc$^{-1}$) & $66.7 \pm 3.4$ & $67.9 \pm 3.1$ & $66.1 \pm 3.0$ \\
$\Omega_m$ & $0.26 \pm 0.06$ & $0.25$ (fixed) & $0.315$ (fixed) \\
$\Xi$ & $0.9550 \pm 0.3710$ & $0.9783 \pm 0.3548$ & $0.9783 \pm 0.3548$ \\
\bottomrule
\end{tabular}
\end{table}

Joint and fixed-$\Omega_m$ inference give nearly the same $\Xi$ posteriors: the median shifts by $\lesssim 2.4\%$ and the width changes by $\lesssim 4.6\%$.
The two fixed-$\Omega_m$ columns therefore show the same $\Xi = 0.9783 \pm 0.3548$ for the same reason as in Table~\ref{tab:xi_om}: at each $\Omega_m$, $H_0$ shifts along the dark-siren distance degeneracy to keep $\Xi\, D_L^{\rm EM}$ fixed, so the marginalized $\Xi$ distribution does not move.
For $\Omega_m$ fixed by a CMB prior, reasonable choices in $[0.20,\,0.35]$ therefore do not bias the $\Xi$ constraint in Table~\ref{tab:xi_om}.
$H_0$ behaves differently.
Joint inference transfers part of the $H_0$--$\Omega_m$ degeneracy into $\Omega_m^{\rm joint}$, lowering the marginalized $H_0$ median by $\sim 1\,{\rm km\,s^{-1}\,Mpc^{-1}}$ relative to the $\Omega_m = 0.25$ fixed case.
Fixed-$\Omega_m$ $H_0$ results must be quoted with the assumed $\Omega_m$; the $\Xi$ constraint is much less sensitive to that choice at O5 sensitivity.

\begin{table}[htbp]
\centering
\caption{Comparison of grid-based and MCMC posteriors at $\Omega_m = 0.25$, obtained with the native \textsc{CHIMERA~2.0} likelihood on the Zenodo O5 mock catalog. Quoted intervals are 68\% credible intervals.}
\label{tab:mcmc}
\begin{tabular}{lccc}
\toprule
Method & $H_0$ (km s$^{-1}$ Mpc$^{-1}$) & $\Xi$ & $|\rho_{H_0 \Xi}|$ \\
\midrule
Grid ($24 \times 24$) & $67.9 \pm 3.1$ & $0.9783 \pm 0.3548$ & $\lesssim 0.05$ \\
MCMC & $69.1 \pm 3.3$ & $1.0030 \pm 0.3370$ & $0.12$ \\
\bottomrule
\end{tabular}
\end{table}

To characterize degeneracies quantitatively, we compute the correlation coefficient $\rho_{H_0 \Xi}$ from the joint two-dimensional posterior at each fixed $\Omega_m$.
We find $|\rho_{H_0 \Xi}| < 0.15$ in all cases, whereas the shift of the $H_0$ median induced by varying $\Omega_m$ at fixed $\Xi$ is much larger than the $H_0$ statistical error.
Full-precision $24 \times 24$ analyses at each adopted $\Omega_m$ confirm that the marginalized $\Xi$ median and 68\% width are invariant across $[0.20,\,0.35]$ to machine precision (both spans $< 10^{-8}$).
The same zero span is recovered at $48 \times 48$ and $96 \times 96$ resolution, so the $\Omega_m$ insensitivity of $\Xi$ is a property of the dark-siren likelihood, not of the $24 \times 24$ discretization.
In the subspace probed by dark sirens, changing the assumed $\Omega_m$ re-centers the $H_0$ posterior along the distance degeneracy without altering the marginalized $\Xi$ distribution.
This conclusion holds when $\Omega_m$ is fixed externally; a full joint inference over $(H_0, \Omega_m, \Xi)$ partially absorbs $\Omega_m$ information into $H_0$ and is treated separately in Sec.~\ref{sec:joint}.

\section{Results}
\label{sec:results}

\subsection{One-dimensional $H_0$ constraint}

\begin{figure}[htbp]
\centering
\includegraphics[width=0.65\textwidth]{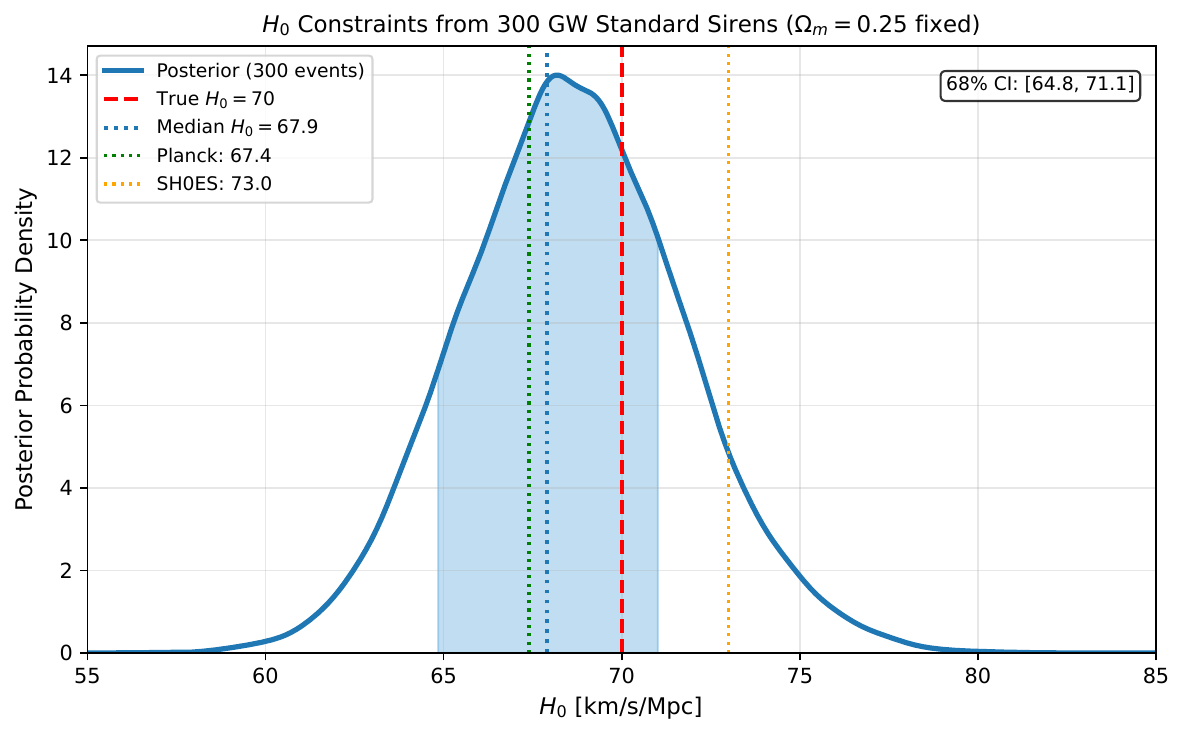}
\caption{Marginalized posterior for $H_0$ from 300 dark sirens with $\Omega_m = 0.25$ fixed. The red dashed line marks the injected value $H_0 = 70$. The posterior peak is $67.9 \pm 3.1\,{\rm km\,s^{-1}\,Mpc^{-1}}$ (68\% credible interval), $2.1\,{\rm km\,s^{-1}\,Mpc^{-1}}$ from the injected value and within the quoted uncertainty.}
\label{fig:h0_1d}
\end{figure}

Figure~\ref{fig:h0_1d} shows the marginalized $H_0$ posterior for 300 events at $\Omega_m = 0.25$, the reference background adopted in the $\Xi$ stability tests.
We obtain
\begin{equation}
H_0 = 67.9 \pm 3.1\,{\rm km\,s^{-1}\,Mpc^{-1}} \qquad (\pm 4.6\%),
\label{eq:h0result}
\end{equation}
consistent with the injected value within $1\sigma$.
This precision matches earlier O5 dark-siren forecasts once differences in detector network, event selection, and population model are taken into account \cite{chen2018,finke2021,leyde2022}.

\subsection{Event-number scaling}

\begin{figure}[htbp]
\centering
\includegraphics[width=\textwidth]{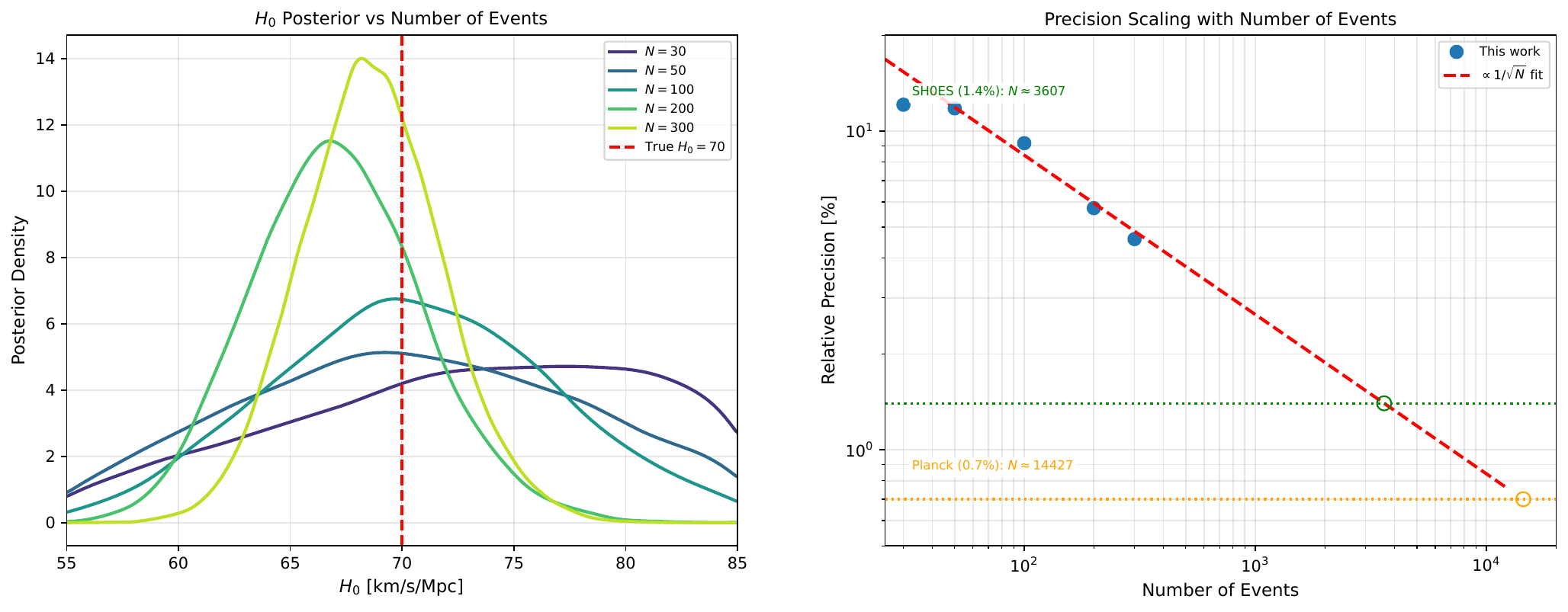}
\caption{Left: marginalized $H_0$ posteriors for subsamples of 30--300 events. Right: relative $H_0$ precision versus event number (blue points) and a $a/\sqrt{N}$ fit with $a = 84.1\% \cdot \sqrt{N}$ (red dashed line). Horizontal lines mark SH0ES ($\sim 1.4\%$) and Planck ($\sim 0.7\%$) precision levels. Intersections of the fit with these lines occur near $N \simeq 3600$ and $N \simeq 1.4 \times 10^4$, respectively; see text for systematic limitations.}
\label{fig:scaling}
\end{figure}

Figure~\ref{fig:scaling} shows progressively sharper $H_0$ posteriors as the sample grows from 30 to 300 events (left panel).
For $N \gtrsim 100$, the relative precision on the right follows the expected $1/\sqrt{N}$ trend.
Fitting the high-$N$ points gives $a = 84.1\% \cdot \sqrt{N}$.
Extrapolating this \textit{statistical} trend to the SH0ES ($1.4\%$) and Planck ($0.7\%$) benchmarks suggests $N \simeq 3600$ and $N \simeq 1.4 \times 10^4$ events, respectively.
These event numbers are statistical extrapolations only.
For dark sirens, $\Omega_m$ degeneracy leaves a systematic floor on $H_0$ at the level of several $\mathrm{km\,s^{-1}\,Mpc^{-1}}$ unless $\Omega_m$ is externally constrained or redshifts are measured (Sec.~\ref{sec:om}).

\subsection{$\Omega_m$ sensitivity and the dominant $H_0$ systematic}
\label{sec:om}

\begin{figure}[htbp]
\centering
\includegraphics[width=\textwidth]{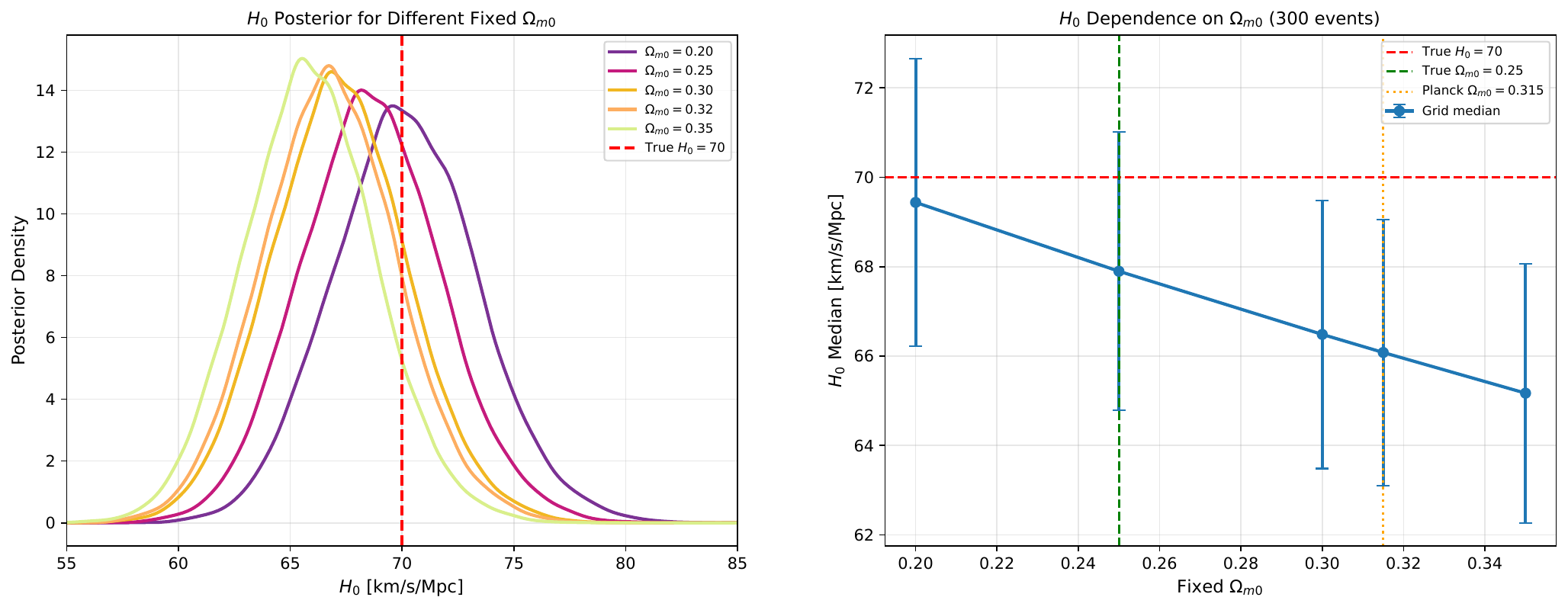}
\caption{Impact of the assumed $\Omega_m$ on the $H_0$ posterior. Left: marginalized $H_0$ distributions for $\Omega_m = 0.20,\,0.25,\,0.30,\,0.315,\,0.35$. Right: $H_0$ median and 68\% interval versus $\Omega_m$. The red star marks the injected cosmology; the green square marks the Planck 2018 value $\Omega_m = 0.315$.}
\label{fig:om_sens}
\end{figure}

Figure~\ref{fig:om_sens} illustrates the $H_0$--$\Omega_m$ degeneracy familiar from dark-siren forecasts \cite{chen2018,finke2021,planck2018}.
Increasing $\Omega_m$ shifts the $H_0$ posterior to lower values without changing its width appreciably.
A linear fit to the median values gives
\begin{equation}
\frac{\mathrm{d}H_0}{\mathrm{d}\Omega_m} \simeq -28\,{\rm km\,s^{-1}\,Mpc^{-1}} \quad {\rm per}\; 1.0\,\Omega_m \;\; (\simeq -2.8\,{\rm km\,s^{-1}\,Mpc^{-1}} \; {\rm per}\; 0.1\,\Omega_m).
\label{eq:om_slope}
\end{equation}
Moving from the injected $\Omega_m = 0.25$ to the Planck value $0.315$ shifts the median from $67.9$ to $66.1\,{\rm km\,s^{-1}\,Mpc^{-1}}$ ($\Delta H_0 \simeq 1.8\,{\rm km\,s^{-1}\,Mpc^{-1}}$), comparable to roughly half the statistical error of a 300-event sample.
By contrast, the marginalized $\Xi$ posterior is unchanged across the same $\Omega_m$ range (Sec.~\ref{sec:xi_stability}): $H_0$ carries the $H_0$--$\Omega_m$ distance degeneracy, not $\Xi$.
This pattern is the leading systematic for dark-siren $H_0$ measurements and can be mitigated by external $\Omega_m$ priors or host redshifts \cite{finke2021,mancarella2022}.

\subsection{Modified-gravity constraint at fixed $\Omega_m$}

\begin{figure}[htbp]
\centering
\includegraphics[width=\textwidth]{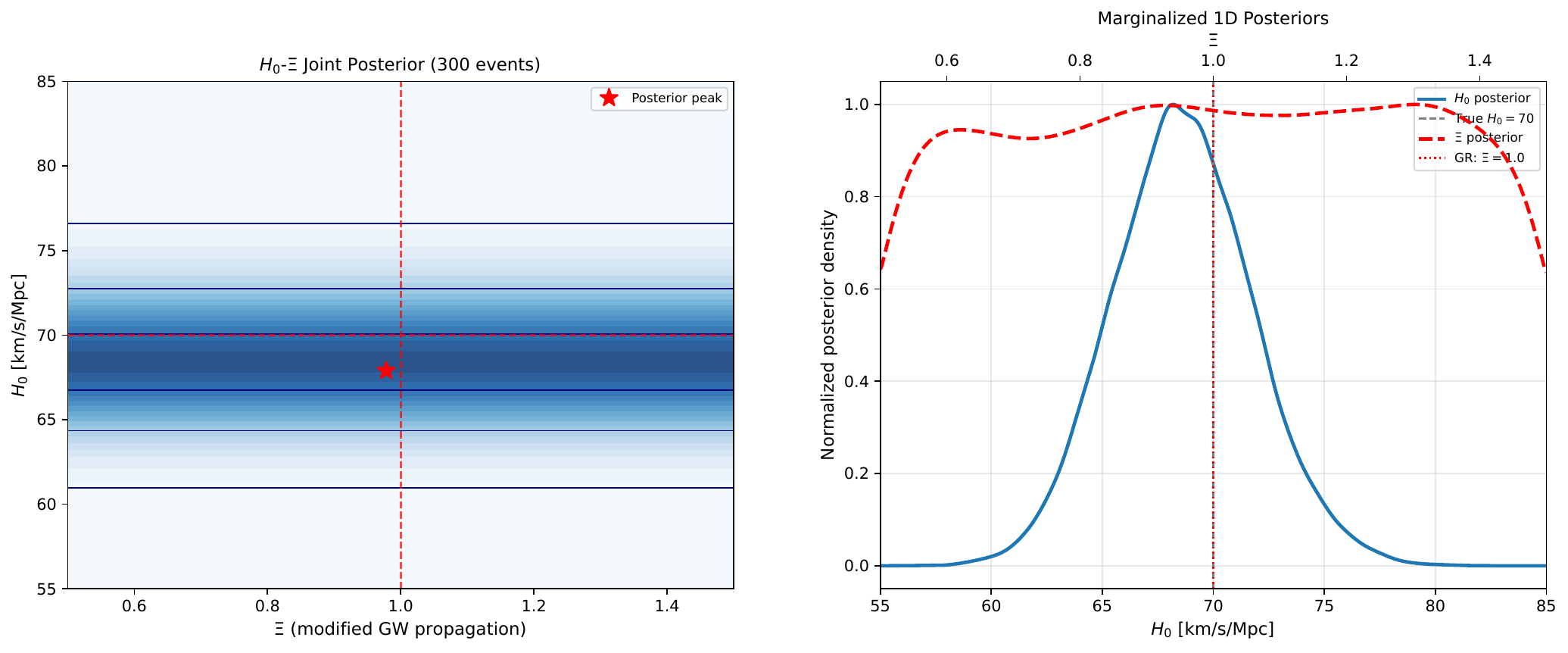}
\caption{Joint $(H_0, \Xi)$ posterior for 300 dark sirens with $\Omega_m = 0.25$. Left: two-dimensional posterior showing a narrow horizontal structure in $H_0$ and a broad direction in $\Xi$. Right: one-dimensional marginals; the $\Xi$ posterior gives $\Xi = 0.9783 \pm 0.3548$ (68\% credible interval), consistent with the GR injection $\Xi = 1.0000$ at current sensitivity.}
\label{fig:xi}
\end{figure}

Figure~\ref{fig:xi} shows the joint $(H_0, \Xi)$ posterior at $\Omega_m = 0.25$.
The two-dimensional distribution is narrow in $H_0$ and much broader in $\Xi$, reflecting the weak individual sensitivity to modified GW propagation without host redshifts.
Without redshift information, $\Xi$ is weakly constrained individually:
\begin{equation}
\Xi = 0.9783 \pm 0.3548 \qquad (\pm 36.3\%),
\label{eq:xiresult}
\end{equation}
consistent with the GR injection $\Xi = 1.0000$ and with the broad constraints reported for related dark-siren forecasts \cite{finke2021,leyde2022}.
The joint posterior has $|\rho_{H_0 \Xi}| < 0.1$, so $H_0$ and $\Xi$ are only weakly correlated at fixed $\Omega_m$.

\subsection{How $\Xi$ responds to variations in $\Omega_m$}
\label{sec:xi_stability}

Table~\ref{tab:xi_om} reports five independent $(H_0,\,\Xi)$ analyses, one at each adopted $\Omega_m$, and constitutes the central robustness test of this work.
The repeated $\Xi$ entries are the forecast outcome, not an absence of new physics: they show quantitatively that the fixed-$\Omega_m = 0.3$ strategy of \cite{tagliazucchi2025} can be extended across CMB-allowed backgrounds without shifting the $\Xi$ constraint, while $H_0$ absorbs the change.

\begin{table}[htbp]
\centering
\caption{Marginalized $(H_0, \Xi)$ constraints for five adopted values of $\Omega_m$. The identical $\Xi$ entries ($0.9783 \pm 0.3548$) and $\Delta\Xi = -0.0217$ reflect the dark-siren distance degeneracy: independent full-precision recomputation at each $\Omega_m$ gives $\Xi = 0.9782609 \pm 0.3547826$ with zero span in both median and width.
At each $\Omega_m$, $H_0$ shifts along the dark-siren distance degeneracy (Eq.~\eqref{eq:om_slope}) to compensate the change in $D_L^{\rm EM}$, while the broad $\Xi$ posterior ($\pm 36.3\%$) and weak $H_0$--$\Xi$ correlation ($|\rho_{H_0 \Xi}| < 0.15$) leave the marginalized $\Xi$ distribution unchanged.
$H_0$ is shown for comparison and moves systematically with $\Omega_m$.
All intervals are 68\% credible intervals from $24 \times 24$ grid-based inference.}
\label{tab:xi_om}
\begin{tabular}{cccc}
\toprule
$\Omega_m$ & $H_0$ (km s$^{-1}$ Mpc$^{-1}$) & $\Xi$ & $\Delta\Xi$ from GR \\
\midrule
0.200 & $69.4 \pm 3.2$ & $0.9783 \pm 0.3548$ & $-0.0217$ \\
0.250 & $67.9 \pm 3.1$ & $0.9783 \pm 0.3548$ & $-0.0217$ \\
0.300 & $66.5 \pm 3.0$ & $0.9783 \pm 0.3548$ & $-0.0217$ \\
0.315 & $66.1 \pm 3.0$ & $0.9783 \pm 0.3548$ & $-0.0217$ \\
0.350 & $65.2 \pm 2.9$ & $0.9783 \pm 0.3548$ & $-0.0217$ \\
\bottomrule
\end{tabular}
\end{table}

Table~\ref{tab:xi_om} lists the main numerical result.
For pure dark sirens, the data constrain the combination $\Xi\, D_L^{\rm EM}(H_0,\,\Omega_m)$: increasing $\Omega_m$ lowers $D_L^{\rm EM}$ at fixed $H_0$, and the posterior compensates by shifting the marginalized $H_0$ median downward by $\simeq 4.2\,{\rm km\,s^{-1}\,Mpc^{-1}}$ from $\Omega_m = 0.20$ to $0.35$ (Eq.~\eqref{eq:om_slope}), preserving the distance scale and hence the marginalized $\Xi$ distribution.
Because $|\rho_{H_0 \Xi}| < 0.15$ and the $\Xi$ posterior is broad ($\pm 36.3\%$), this compensation is absorbed almost entirely by $H_0$.
Full-precision recomputation finds no variation in the $\Xi$ median, credible-interval width, or $\Delta\Xi = -0.0217$ offset from GR across $[0.20,\,0.35]$; the repeated table entries are therefore a direct consequence of the $H_0$--$\Omega_m$ distance degeneracy, in agreement with Eq.~\eqref{eq:jointresult}.

Earlier dark-siren modified-gravity forecasts \cite{finke2021,leyde2022,mancarella2022} reported broad $\Xi$ posteriors but did not explore the $\Omega_m$ dependence of $\Xi$ systematically within the \textsc{CHIMERA~2.0} O5 mock set.
\cite{tagliazucchi2025} adopted a single fixed $\Omega_m = 0.3$ with galaxy catalogs and likewise did not report how the pure dark-siren $\Xi$ posterior changes under alternative $\Omega_m$ assumptions.

\subsection{Comparison with the galaxy-catalog scenario}

\begin{table}[htbp]
\centering
\caption{Precision comparison between the present pure dark-siren study of varying $\Omega_m$ and the galaxy-catalog forecast of \cite{tagliazucchi2025} with fixed $\Omega_m = 0.3$, based on the same 300 O5 events. Gains are defined as the ratio of quoted percentage uncertainties.}
\label{tab:comparison}
\begin{tabular}{cccc}
\toprule
Parameter & This work & Tagliazucchi \textit{et al.} 2025 & Gain from redshifts \\
\midrule
$H_0$ & $\pm 4.6\%$ & $\sim \pm 2\%$ & $\sim 2\times$ \\
$\Xi$ & $\pm 36.3\%$ & $\sim \pm 7.5\%$ & $\sim 5\times$ \\
\bottomrule
\end{tabular}
\end{table}

Table~\ref{tab:comparison} shows that spectroscopic hosts improve $\Xi$ precision substantially more than $H_0$ precision.
This comparison isolates the main practical difference from \cite{tagliazucchi2025}: on the same 300 events, absent host redshifts widen the $\Xi$ posterior to $\pm 36.3\%$, whereas their galaxy-catalog analysis reaches $\sim 7.5\%$ at fixed $\Omega_m = 0.3$.
The bottleneck for percent-level modified-gravity tests is therefore redshift information, not the choice among CMB-allowed $\Omega_m$ values.
Without measured $z$, the combination $\Xi\, D_L^{\rm EM}(H_0,\,\Omega_m)$ is underdetermined event by event.
When $\Omega_m$ is varied at fixed data, $H_0$ moves along the distance degeneracy to keep this combination fixed, leaving the marginalized $\Xi$ posterior unchanged; stacking many events sharpens $H_0$ through the population distance distribution but does not break the $\Xi$--background confusion \cite{finke2021,laghi2021}.

\section{Discussion}
\label{sec:discussion}

Our $H_0$ precision at 300 events agrees with earlier O5 dark-siren forecasts \cite{chen2018,finke2021,leyde2022}, as does the broad $\Xi$ posterior found without host redshifts \cite{finke2021,leyde2022,mancarella2022}.
The additional step taken here is to vary $\Omega_m$ over $[0.20,\,0.35]$ in the \textsc{CHIMERA~2.0} O5 mocks while keeping the pure dark-siren setup.
Across that interval we find $\Xi = 0.9783 \pm 0.3548$, with $H_0$ shifting by up to $\sim 4.2\,{\rm km\,s^{-1}\,Mpc^{-1}}$.
The same behavior appears in the joint $(H_0,\,\Omega_m,\,\Xi)$ analysis (Table~\ref{tab:joint}) and in the grid/MCMC comparison at $\Omega_m = 0.25$ (Table~\ref{tab:mcmc}), and follows the familiar dark-siren distance degeneracy discussed in Secs.~\ref{sec:om} and \ref{sec:xi_stability}.

For O5 pipeline design, a CMB-informed fixed $\Omega_m$ is unlikely to add a further $\Xi$ systematic at the precision reached here.
The comparison in Table~\ref{tab:comparison} shows that the main penalty relative to galaxy-catalog forecasts on the same catalog is the absence of measured host redshifts.
Dark sirens alone are unlikely to reach percent-level $\Xi$ sensitivity, and host follow-up remains necessary for tighter modified-gravity tests \cite{belgacem2018,laghi2021,tagliazucchi2025}.

\label{sec:limitations}
The analysis has several limitations.
Population hyperparameters are fixed to the injected values, and population-model uncertainties can bias cosmological inference at levels comparable to the statistical errors of small samples \cite{abbott2022population,mandel2019}.
Varying $\gamma$, $\kappa$, or the mass model would be a natural extension.
The mock catalog contains only BBHs; adding neutron-star binaries would change both the counterpart fraction and the redshift distribution.
We kept the dark-energy equation of state and $\Xi(z)$ models fixed \cite{belgacem2019,ezquiaga2018}.
The extrapolation to $N \simeq 1.4 \times 10^4$ events reaches Planck-like $H_0$ precision only statistically; the $\Omega_m$ systematic remains important for dark-siren $H_0$.
Einstein Telescope and Cosmic Explorer may detect $\sim 10^5$ events per year \cite{maggiore2020}.
At that event rate, the $H_0$--$\Omega_m$ degeneracy found here will dominate the $H_0$ error budget, whereas percent-level $\Xi$ measurements will require both dark sirens and deep galaxy catalogs.

\section{Conclusions}
\label{sec:conclusions}

We analyzed 300 pure dark sirens in the public \textsc{CHIMERA~2.0} O5 mock catalog with $\Omega_m$ varied between $0.20$ and $0.35$.
Without host redshifts, the marginalized $\Xi$ posterior is $0.9783 \pm 0.3548$ ($\pm 36.3\%$) and is independent of the adopted background over that interval (Table~\ref{tab:xi_om}).
The data constrain $\Xi\, D_L^{\rm EM}(H_0,\,\Omega_m)$, so a change in $\Omega_m$ is absorbed by $H_0$ (Eq.~\eqref{eq:om_slope}) and the $\Xi$ marginal is unchanged.
Joint $(H_0,\,\Omega_m,\,\Xi)$ inference gives $\Xi^{\rm joint} = 0.9550 \pm 0.3710$ and $|\rho_{\Omega_m \Xi}| \lesssim 0.05$.
$H_0$ remains strongly background dependent ($67.9 \pm 3.1\,{\rm km\,s^{-1}\,Mpc^{-1}}$ at $\Omega_m = 0.25$, $66.7 \pm 3.4\,{\rm km\,s^{-1}\,Mpc^{-1}}$ in the joint fit, and a median shift of $\simeq 4.2\,{\rm km\,s^{-1}\,Mpc^{-1}}$ from $\Omega_m = 0.20$ to $0.35$) and should always be reported with the assumed or inferred $\Omega_m$.
On the same events, the main obstacle to percent-level $\Xi$ constraints is the lack of host redshifts (Table~\ref{tab:comparison}: $\pm 36.3\%$ versus $\sim 7.5\%$ with galaxy catalogs), not the choice of $\Omega_m$ within the CMB-allowed range.

\acknowledgments

This work is funded by the National Astronomical Observatories of the Chinese Academy of Sciences, Project No.~E4TG6601, and has been supported in part by the National Key Research and Development Program of China under Grant No.~2021YFC2203000.

\bibliographystyle{JHEP}
\bibliography{references}

\providecommand{\href}[2]{#2}\begingroup\raggedright\begin{thebibliography}{10}

\bibitem{planck2018}
{Planck Collaboration}, \emph{Planck 2018 results. {VI}. cosmological
  parameters}, \href{https://doi.org/10.1051/0004-6361/202039196}{\emph{Astron.
  Astrophys.} {\bfseries 641} (2020) A6}
  [\href{https://arxiv.org/abs/1807.06209}{{\ttfamily 1807.06209}}].

\bibitem{riess2022}
A.G.~Riess et~al., \emph{A comprehensive measurement of the local value of the
  hubble constant with 1 km s$^{-1}$ mpc$^{-1}$ uncertainty from the hubble
  space telescope and the {SH0ES} team},
  \href{https://doi.org/10.3847/1538-4357/ac5c5b}{\emph{Astrophys. J.}
  {\bfseries 934} (2022) L7}
  [\href{https://arxiv.org/abs/2112.04510}{{\ttfamily 2112.04510}}].

\bibitem{schutz1986}
B.F.~Schutz, \emph{Determining the hubble constant from gravitational wave
  observations}, \href{https://doi.org/10.1038/323310a0}{\emph{Nature}
  {\bfseries 323} (1986) 310}.

\bibitem{holz2005}
D.E.~Holz and S.A.~Hughes, \emph{Using gravitational-wave standard sirens},
  \href{https://doi.org/10.1086/431341}{\emph{Astrophys. J.} {\bfseries 629}
  (2005) 15} [\href{https://arxiv.org/abs/astro-ph/0508116}{{\ttfamily
  astro-ph/0508116}}].

\bibitem{gw170817}
{LIGO Scientific and Virgo Collaborations}, \emph{A gravitational-wave standard
  siren measurement of the hubble constant},
  \href{https://doi.org/10.1038/nature24471}{\emph{Nature} {\bfseries 551}
  (2017) 85} [\href{https://arxiv.org/abs/1710.05835}{{\ttfamily 1710.05835}}].

\bibitem{abbott2020prospects}
B.P.~Abbott et~al., \emph{Prospects for observing and localizing
  gravitational-wave transients with {Advanced LIGO}, {Advanced Virgo} and
  {KAGRA}}, \href{https://doi.org/10.1007/s41114-020-00026-9}{\emph{Living Rev.
  Relativ.} {\bfseries 23} (2020) 3}
  [\href{https://arxiv.org/abs/1304.0670}{{\ttfamily 1304.0670}}].

\bibitem{belgacem2018}
E.~Belgacem, Y.~Dirian, S.~Foffa and M.~Maggiore, \emph{Modified
  gravitational-wave propagation and standard sirens},
  \href{https://doi.org/10.1088/1475-7516/2018/07/024}{\emph{JCAP} {\bfseries
  07} (2018) 024} [\href{https://arxiv.org/abs/1806.08794}{{\ttfamily
  1806.08794}}].

\bibitem{borghi2024}
N.~Borghi, M.~Mancarella, M.~Moresco, M.~Tagliazucchi, F.~Iacovelli, A.~Cimatti
  et~al., \emph{Cosmology and astrophysics with standard sirens and galaxy
  catalogs in view of future gravitational wave observations},
  \href{https://doi.org/10.3847/1538-4357/ad20eb}{\emph{Astrophys. J.}
  {\bfseries 964} (2024) 191}
  [\href{https://arxiv.org/abs/2312.05302}{{\ttfamily 2312.05302}}].

\bibitem{tagliazucchi2025}
M.~Tagliazucchi, M.~Moresco, N.~Borghi and M.~Fiebig, \emph{Accelerating the
  standard siren method: Improved constraints on modified gravitational-wave
  propagation with future data},
  \href{https://doi.org/10.1051/0004-6361/202554827}{\emph{Astron. Astrophys.}
  {\bfseries 702} (2025) A244}
  [\href{https://arxiv.org/abs/2504.02034}{{\ttfamily 2504.02034}}].

\bibitem{fishbach2020fossil}
M.~Fishbach and D.E.~Holz, \emph{Fossil colliding compact objects as dark
  sirens to measure the {Hubble} parameter},
  \href{https://doi.org/10.3847/2041-8213/ab7a6a}{\emph{Astrophys. J. Lett.}
  {\bfseries 891} (2020) L31}
  [\href{https://arxiv.org/abs/2002.07656}{{\ttfamily 2002.07656}}].

\bibitem{finke2021}
A.~Finke, S.~Foffa, F.~Iacovelli, M.~Maggiore and M.~Mancarella,
  \emph{Cosmology with {LIGO}/{Virgo} dark sirens: Hubble parameter and
  modified gravitational wave propagation}, {\emph{JCAP} {\bfseries 08} (2021)
  026} [\href{https://arxiv.org/abs/2101.09069}{{\ttfamily 2101.09069}}].

\bibitem{leyde2022}
K.~Leyde, S.~Mastrogiovanni, D.A.~Steer, E.~Chassande-Mottin and
  C.~Karathanasis, \emph{Current and future constraints on cosmology and
  modified gravitational wave friction from binary black holes}, {\emph{JCAP}
  {\bfseries 09} (2022) 012}
  [\href{https://arxiv.org/abs/2204.11998}{{\ttfamily 2204.11998}}].

\bibitem{mancarella2022}
M.~Mancarella, E.~Genoud-Prachex and M.~Maggiore, \emph{Cosmology and modified
  gravity with dark sirens from {GWTC-3}}, {\emph{Phys. Rev. D} {\bfseries 105}
  (2022) 064030} [\href{https://arxiv.org/abs/2110.14805}{{\ttfamily
  2110.14805}}].

\bibitem{chimera2025zenodo}
M.~Tagliazucchi, M.~Moresco and N.~Borghi, \emph{{CHIMERA}: v2.0},  2025.
\newblock 10.5281/zenodo.17143346.

\bibitem{madau1998}
P.~Madau, L.~Pozzetti and M.~Dickinson, \emph{The star formation history of
  field galaxies}, \href{https://doi.org/10.1086/305523}{\emph{Astrophys. J.}
  {\bfseries 498} (1998) 106}.

\bibitem{chen2018}
H.-Y.~Chen, M.~Fishbach and D.E.~Holz, \emph{A two per cent hubble constant
  measurement from standard sirens within five years}, {\emph{Nature}
  {\bfseries 562} (2018) 545}
  [\href{https://arxiv.org/abs/1712.06531}{{\ttfamily 1712.06531}}].

\bibitem{laghi2021}
D.~Laghi, N.~Tamanini, W.~Del~Pozzo and A.~Sesana, \emph{Testing modified
  gravity with gravitational wave standard sirens: Challenges and prospects},
  {\emph{Mon. Not. Roy. Astron. Soc.} {\bfseries 508} (2021) 4512}
  [\href{https://arxiv.org/abs/2103.01105}{{\ttfamily 2103.01105}}].

\bibitem{abbott2022population}
R.~Abbott et~al., \emph{Population of merging compact binaries inferred using
  gravitational waves through {GWTC-3}},
  \href{https://doi.org/10.1103/PhysRevX.13.011048}{\emph{Phys. Rev. X}
  {\bfseries 13} (2022) 011048}
  [\href{https://arxiv.org/abs/2111.03634}{{\ttfamily 2111.03634}}].

\bibitem{mandel2019}
I.~Mandel and W.M.~Farr, \emph{Extracting distribution parameters from multiple
  uncertain observations with selection biases}, {\emph{Mon. Not. Roy. Astron.
  Soc.} {\bfseries 486} (2019) 1086}
  [\href{https://arxiv.org/abs/1901.08877}{{\ttfamily 1901.08877}}].

\bibitem{belgacem2019}
E.~Belgacem et~al., \emph{Testing modified gravity at cosmological distances
  with {LISA} standard sirens}, {\emph{JCAP} {\bfseries 07} (2019) 024}
  [\href{https://arxiv.org/abs/1903.08618}{{\ttfamily 1903.08618}}].

\bibitem{ezquiaga2018}
J.M.~Ezquiaga and M.~Zumalac\'{a}rregui, \emph{Dark energy in light of
  multi-messenger gravitational-wave astronomy}, {\emph{Front. Astron. Space
  Sci.} {\bfseries 5} (2018) 44}
  [\href{https://arxiv.org/abs/1807.09282}{{\ttfamily 1807.09282}}].

\bibitem{maggiore2020}
M.~Maggiore et~al., \emph{Science case for the einstein telescope},
  {\emph{JCAP} {\bfseries 03} (2020) 050}
  [\href{https://arxiv.org/abs/1907.04833}{{\ttfamily 1907.04833}}].

\end{thebibliography}\endgroup

\end{document}